\documentclass[twocolumn]{aastex63}

\usepackage{commath}
\usepackage{epsfig}

\newcommand{\Ms}{{\ensuremath{\rm M_{\odot} }}}

\submitjournal{ApJ}

\shorttitle{Pop III stars }
\shortauthors{Latif  et al.}

\begin{document}

\title{The Birth Mass Function of Pop III Stars}



\correspondingauthor{Muhammad A. Latif}
\email{latifne@gmail.com}

\author{Muhammad A. Latif}
\affiliation{Physics Department, College of Science, United Arab Emirates University, PO Box 15551, Al-Ain, UAE}

\author{Daniel Whalen}
\affiliation{Institute of Cosmology and Gravitation, University of Portsmouth, Portsmouth PO1 3FX, UK}

\author{Sadegh Khochfar}
\affiliation{Institute for Astronomy, University of Edinburgh, Royal Observatory, Blackford Hill, Edinburgh EH9 3HJ, UK}

\begin{abstract}

Population III (Pop III) stars ended the cosmic Dark Ages and began early cosmological reionization and chemical enrichment.  However, in spite of their importance to the evolution of the early Universe, their properties remain uncertain because of limitations to previous numerical simulations and the lack of any observational constraints. Here we investigate Pop III star formation in five primordial halos with 3D radiation-hydrodynamical cosmological simulations. We find that multiple stars form in each minihalo and that their numbers increase over time, with up to 23 stars forming in one of the halos.  Radiative feedback from the stars generates strong outflows, deforms the surrounding protostellar disk, and delays star formation for a few thousand years.  Star formation rates vary with halo and depend on mass accretion onto the disk, halo spin number, and the fraction of massive stars in the halo. Stellar masses in our models range from  0.1-37 \Ms, and of the 55 stars that form in our models twelve are $\rm > 10~ \Ms$ and most of the others are 1-10 \Ms.  Our simulations thus suggest that Pop III stars have characteristic masses of 1-10 \Ms\ and a top-heavy IMF with dN/dM $\propto M_*^{-1.18}$. Up to 70\% of  the stars are ejected from their disks by three-body interactions which, along with ionizing UV feedback, limits their final masses.

\end{abstract}


\section{Introduction} \label{sec:intro}

The emergence of Population III (Pop III) stars 100 - 200 million years after the Big Bang ended the cosmic Dark Ages and began cosmic reionization, early chemical enrichment enrichment, and the formation of the first stellar black holes.  Pop III stars are expected to form in $10^5 - 10^6$ \Ms\ primordial minihalos at $z= 20-30$ as a result of gravitational collapse induced by $\rm H_2$ ro-vibrational line cooling in the absence of metals. Their initial mass function (IMF) is still an open question and is central to our understanding of the high-redshift universe.

The first studies suggested that Pop III stars have typical masses of a few hundred solar masses and  form in isolation, one per halo \citep{Bromm02,Abel02,Yoshida08}. However, recent numerical simulations show that the accretion disk at the center of the halo fragments into multiple clumps that can produce multiple stars \citep{Clark11,Latif13ApJ,Hirano14,Sharda20}. \cite{Greif12} found that almost two thirds of fragments later merged and only a third survived 10 years after the formation of the protostar.  \cite{hir15} considered only single stars forming per minihalo and found that their masses range from 10 - 1000 \Ms. \cite{Stacy16} performed a cosmological simulation with feedback from the most massive protostar only and found that 37 sink particles form in one minihalo and the most massive star grows to 20 \Ms\ in the first 5 kyr. Radiation hydrodynamical simulations by \cite{Hosokawa16} in which only one star was allowed to form in the disk found that the star  grows to 600 \Ms\ by clump migration \citep[see also][]{Latif15a}. However, it is not clear how radiation from multiple stars in the disk would affect the outcome of these studies.

\cite{Susa19} followed the collapse of a Bonner-Ebert sphere with a simple barotropic equation of state without stellar feedback or chemistry and found that the number of fragments in the disk increase with time after the onset of star formation ($ \rm \propto t^{0.3}$).  More recently, \cite{Sugi20} simulated the collapse of a cloud extracted from a cosmological simulation at densities of $\rm 10^6~cm^{-3}$ with six-species chemistry and radiative feedback from protostars and found that it formed a massive, wide binary system (60 \Ms\ and 70 \Ms).  These studies either either included feedback from just one protostar in a cosmological environment or relied on idealized initial conditions to study fragmentation in primordial gas clouds. No cosmological simulation of Pop III star formation has ever included ionizing and dissociative feedback from all the protostars in the cloud.

We have performed 3D radiation-hydrodynamical cosmological simulations of Pop III star formation in five primordial halos at $\rm z=20-30$ with sink particles that mimic Pop III stars and mass-dependent photodissociative and ionizing feedback from the protostars.  The simulations are evolved for up to $\rm \sim 20 ~kyr$ after the formation of the first star.  We summarize our simulation setup and recipe for star formation and feedback in Section 2. Our main results are presented in Section 3 and we conclude in Section 4.

\section{Numerical Method} \label{sec:methods}

We use the Enzo adaptive mesh refinement (AMR) code \citep{Enzo14} for the simulations in our study. They are initialized at $z=150$ with Gaussian primordial density fluctuations generated by MUSIC \citep{Hahn11} with cosmological parameters from the Planck 2016 data release \citep{Planck16}.  The simulation box is 300 $h^{-1}$ ckpc on a side with a top grid resolution of $256^3$ and two additional static nested grids centered on the halo for an effective initial resolution of $\rm 1024^3$.   We allow up to 20 levels of refinement in the subvolume covering 20\% of the top grid, which produce an effective spatial resolution of $\rm \sim 10$ AU.  DM particles are split into 13 child particles, which yields an effective DM resolution of about 0.2 \Ms. The Jeans length throughout the simulation is resolved by at least 32 cells. For further details about our refinement criteria see \cite{Latif21a}.  

\subsection{Sink Formation / Stellar Feedback}

We use sink particles to represent Pop III stars.  Sinks are created in a grid cell that meet the following conditions \citep{Regan18b,Latif21a}:  (1) it is at the maximum refinement level, (2) it is at the local minimum of the gravitational potential, (3) it has a convergent flow, (4) the gas density is higher than the Jeans density, and (5) the cooling time is shorter than the free-fall time.  These criteria typically create sinks at densities $\rm \geq 10^{-12}~g/cm^3$.  A sink particle can accrete gas from a radius of 4 cells, and any other particles forming within the accretion radius are immediately merged with the most massive one.  Sink particles are also merged with the more massive sink if they come within each other's accretion radius.  The velocity of the sink after accretion or a merger is determined from conservation of momentum \citep{Krumholz004}.  The accretion rate of the sink is calculated from the mass influx at the accretion radius. \citet{Fed10} include the boundedness of the gas and the Jeans instability in their criteria but \citet{Regan18b} found that their effects on sink formation are less important than those of the criteria listed above.

Radiative feedback from a protostar depends on its effective temperature and luminosity, which can be determined from its accretion history, mass, and radius. Pop III prototstars are usually born on the Hayashi track with effective temperatures of $\rm \sim $ 5000 K and later transition to the Henyey track before eventually reaching the main sequence \cite[see Section 2 of][and references therein, hereafter S16]{Stacy16}.  The radius of the protostar depends on whether accretion proceeds through a thin disk or is spherical \citep{Hos10}.  In either case, the star begins Kelvin-Helmholtz contraction at 
\begin{equation}
\rm M_{*}  \sim 7  \, \left( \frac{\dot{M}} {10^{-3}~\Ms/yr} \right)^{0.27}  \Ms,
\vspace{-0.05in}
\end{equation}
where $\dot{M}$ is the accretion rate of the protostar \citep{Omukai03,Hos10,Smith12}.  S16 found that the star only produces ionizing UV after growing to $\gtrsim$ 10 \Ms.  We assume here that protostars remain cool and do not produce any ionizing radiation until they reach 10 \Ms, when they begin to emit both IR and ionizing UV.  The exact transition to the zero-age main sequence is uncertain and depends on the geometry of accretion (disk or spherical) and the accretion rates of the protostars. S16 compared several models for this transition and found that they only begin to converge at $\rm > 5$ \Ms. Furthermore, disk accretion models exhibit an abrupt increase in stellar luminosity above a few solar masses (see right panel of Fig. 1 in Stacy et al 2017).  Our choice of switching on ionizing UV at 10 $\rm \Ms$ is thus consistent with previous studies \citep{Omukai03,Hos10,Smith12,Stacy16}. 

We therefore set $\rm T_{eff}  = $ 5000 K when $\rm M_{*}  \leq$ 10 \Ms\ \citep{Hos13} and $\rm T_{eff} = 10^{4.759}$ when $\rm M_{*} >$ 10 \Ms\ \citep{Schaerer02}, and assume that the protostellar mass is equal to the sink mass.  \citet{Hosokawa16} found that short bursts of rapid accretion can lead to protostellar expansion even above 10 \Ms.  To account for such episodes, we assume that $\rm T_{eff} =$ 5000 K above a threshold accretion rate of $\sim$ 10$^{-2}$ \Ms\ yr$^{-1}$ and $\rm T_{eff} = 10^{4.759}$ below this rate.  Stars above 10 \Ms\ are approximated as blackbodies with $\rm T_{eff} = 10^{4.759}$.  The luminosity of the star scales with mass but $\rm T_{eff}$ is held constant for simplicity because it does not change much over the masses of the stars in our study \citep[see Table 3 of][]{Schaerer02}.  This approximation has little effect on the dynamics of the I-fronts of the stars because ionized gas temperatures do not vary strongly with photon energy, or hence $\rm T_{eff}$ \citep{Whalen08}.  

We use the MORAY ray tracing radiation transport module \citep{Wise11} to propagate radiation from the stars and partition the flux into four energy bins: one for photodetachment of H$^{-}$ (2.0 eV), one for Lyman-Werner photodissociation of H$_{2}$ and H$_{2}^{+}$ (12.8 eV), and two for ionizations of H and He (14.0 eV, 25.0 eV).  Energy fractions of 0.3261, 0.1073, 0.3686, 0.1965 are used in bins 1 - 4, respectively, and are taken from Table 4 of \citet{Schaerer02}.  Stars below 10 \Ms\ are assumed to only be sources of IR peaking at 2 eV.  Radiation pressure due to momentum transfer by ionizations is included in MORAY \citep{Wise12a}.

\subsection{Primordial Gas Chemistry}

We use the non-equilibrium primordial gas reaction network from \citet{Turk12}, which is based on \citet{Abel97} and \citet{Anni97}, to evolve H, H$^+$, H$^-$, He, He$^+$, He$^{2+}$, H$_2$, H$_2^+$, and e$^-$ mass fractions.  Collisional ionization and excitation cooling by H and He, recombination cooling, H$_2$ cooling, inverse Compton (IC) cooling, bremsstrahlung cooling, collisionally-induced emission (CIE) cooling at high densities, and heating due to three-body reactions are all included in updates to the gas energy equation.  At densities above 10$^{-14}$ g cm$^{-3}$ the gas becomes optically-thick to $\rm H_2 $ lines so we reduce the optically-thin H$_2$ cooling rate by fitting factors from \citet{Ripa04}. We use the \citet{Glover08} rate coefficients for three-body H$_2$ formation and the \cite{WG11} model for $\rm H_2$ self-shielding from LW radiation.  We do not include deuterium or related species because HD cooling mostly occurs in relic H II regions or shock-heated gas during major mergers, which can boost D$^+$ abundances \citep{McG08,Greif08,Bov14}.  Our chemistry solver is self-consistently coupled to hydrodynamics and radiation transport.

\section{Results} \label{sec:results}

\begin{table}
\begin{center}
\begin{tabular}{| c | c | c | c| c|}
\hline
\hline
Halo & $z$  & Mass  & $\rm \lambda$ & Stellar Mass  \\
    &         &            (\Ms)            &                  &  (\Ms) \\
\hline
1  &  22  &  $\rm 2.7 \times 10^5$  &  0.042  & 166.02 \\
2  &  28  &  $\rm 1.8 \times 10^5$  &  0.011  &  83.64  \\ 
3  &  20  &  $\rm 3.2 \times 10^5$  &  0.016  & 31.57\\
4  &  21  &  $\rm 2.7 \times 10^5$  &  0.025  & 27.10\\
5  &  22  &  $\rm 5.2 \times 10^5$  &  0.025  & 56.59 \\
\hline
\end{tabular}
\caption{Virial masses, collapse redshifts, spin parameters $\lambda$, and total mass of Pop III stars of the halos in our study.}
\label{tbl:tbl1}
\end{center}
\end{table}

\begin{figure*}
\centering
\includegraphics[width=0.9\textwidth]{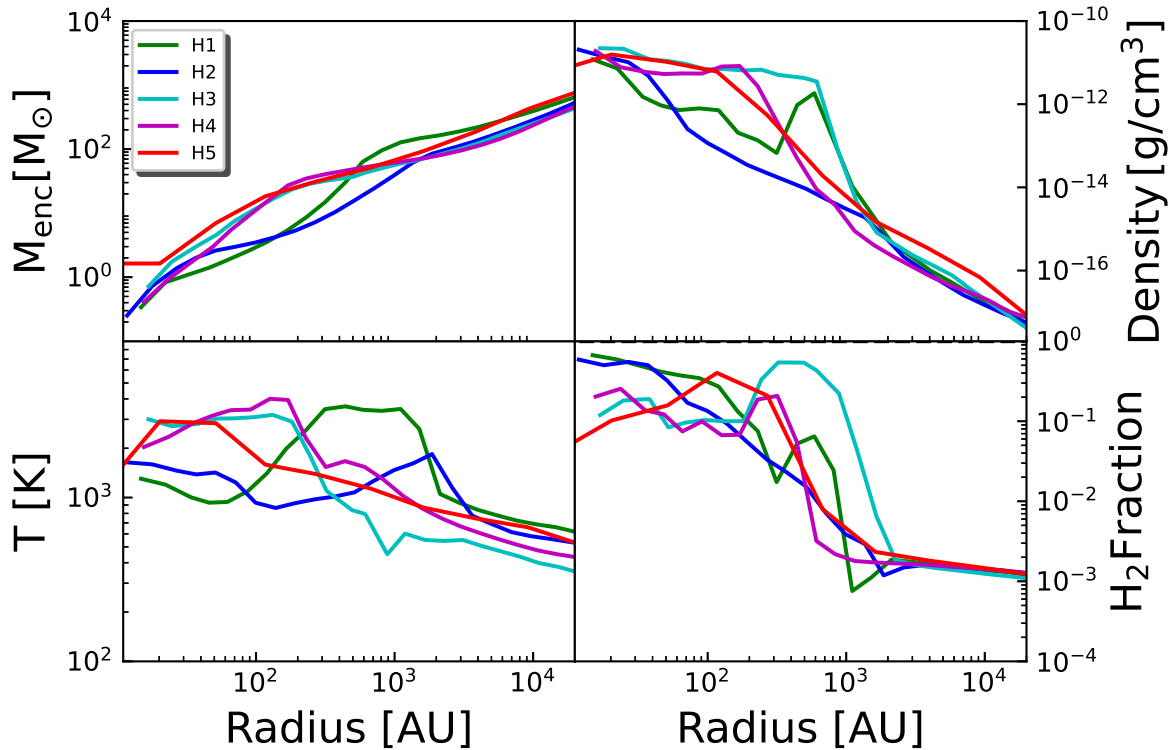}
\caption{Spherically-averaged radial profiles of gas density, temperature, H$_2$ mass fraction and enclosed mass at the onset of star formation of the five halos.}
\label{fig:f1}
\end{figure*}

We simulated Pop III star formation in five halos whose masses and collapse redshifts are listed in Table~\ref{tbl:tbl1}. They are evolved from different Gaussian random fields and ave spin parameters that are sampled from the peak in spin distribution of minihalos from large-scale numerical simulations \citep{Bul01}. Twenty levels of refinement allow us to follow the collapse of gas in these halos from kpc scales down to about 10 AU. The trace amounts of $\rm H_2$ formed in gas phase reactions are boosted during virialization and trigger collapse in halos above a few 10$^5$ \Ms.  Collapse leads to the formation of an accretion disk in each halo with an initial mass of $\sim$ 100 \Ms\ and radius of a few hundred AU.   In Figure~\ref{fig:f1} we show spherically-averaged profiles of density, temperature, H$_2$ mass fraction and enclosed mass in the center of each halo at the onset of star formation in the disks.

Densities rise from $\sim$ 10$^{-17}$ g cm$^{-3}$ to 10$^{-11}$ g cm$^{-3}$ from 20,000 AU down to 10 AU.  Above 10$^{-16}$ g cm$^{-3}$, three-body reactions boost H$_2$ mass fractions from $\sim$ 10$^{-3}$ to as high as 0.5 at the center of the disk.  $\rm H_2$ initially cools gas to a few hundred K but at the high densities closer to the center of the disk the gas becomes opaque to $\rm H_2$ lines which, together with heating due to three-body formation of H$_2$, raises temperatures to a few thousand K.  H$_2$ fractions in the disks vary from 0.2 - 0.4 at their centers to a few 10$^{-1}$ at their outer edges.

\begin{figure}
\centering
\includegraphics[width=0.5\textwidth]{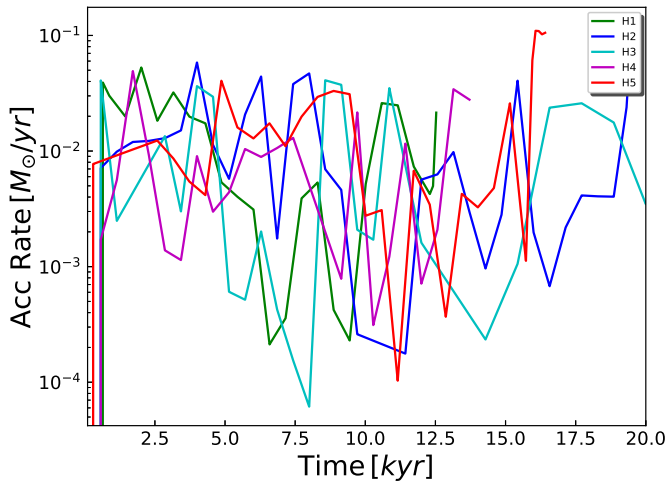}
\caption{Accretion rates for the disk (central 300 AU) in each halo.}
\label{fig:f2}
\end{figure}

Small differences in temperatures and $\rm H_2$ abundances between disks are due to variations in densities between the halos.  As shown in Figure~\ref{fig:f2}, accretion rates onto the disks fluctuate between 10$^{-4}$ \Ms\ yr$^{-1}$ and 0.05 \Ms\ yr$^{-1}$ but average $\sim$ 0.01 \Ms\ yr$^{-1}$ except in halo 2, which has about half this rate. These rates are a factor of a few lower than estimates from $\rm M_{Jeans}/T_{ff} \propto T^{3/2}$, which suggests that accretion is regulated by radiation from stars.

\begin{figure*} 
\begin{center}
\includegraphics[scale=0.8]{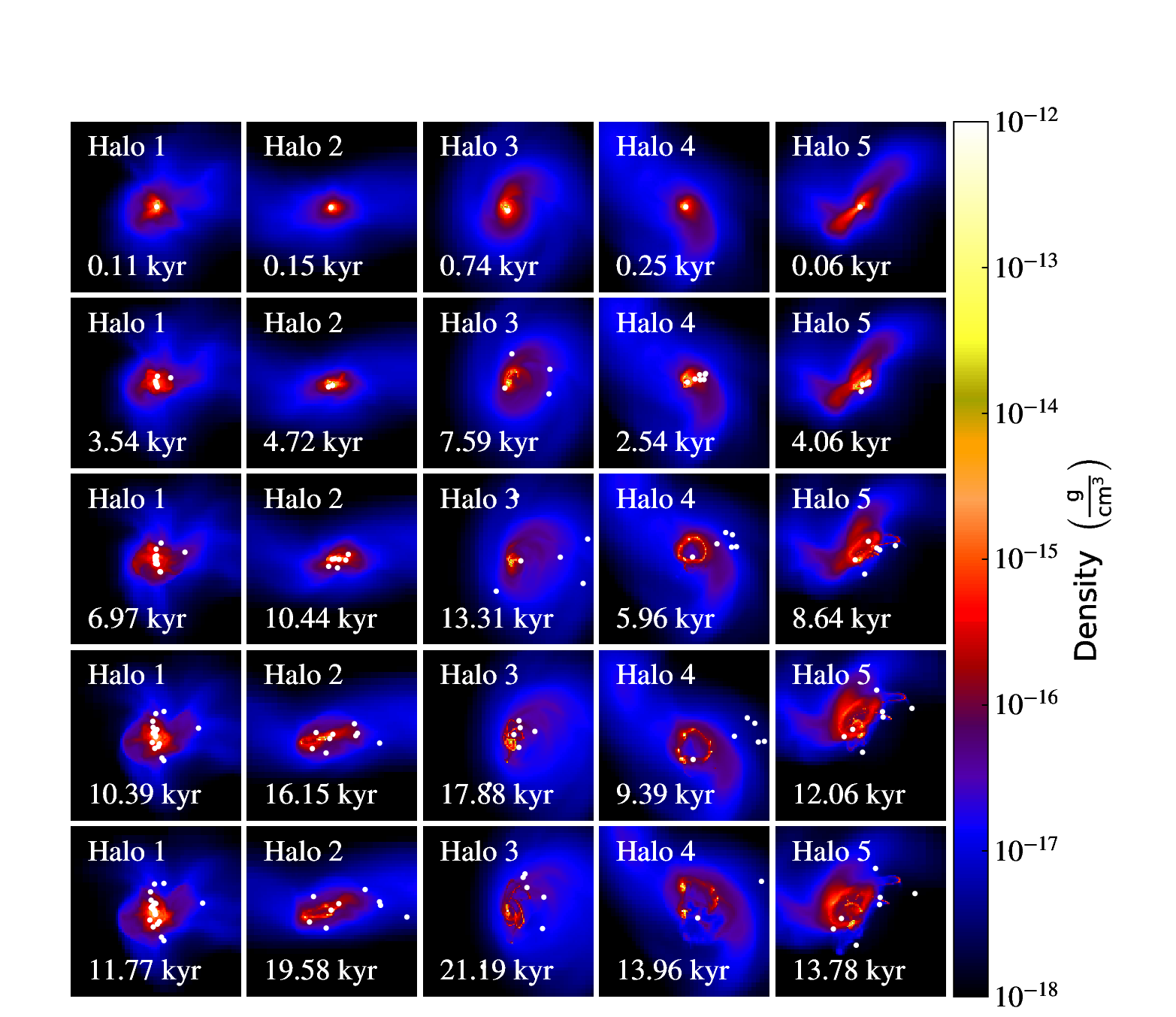}
\end{center}
\caption{Density projections of the central 0.2 pc of each halo. Each column shows the time evolution of the halo from top to bottom, times are measured from the formation of the first star in the halo.  White dots mark the locations of star particles.}
\label{fig:f3}
\end{figure*}

\begin{figure*} 
\begin{center}
\includegraphics[scale=0.8]{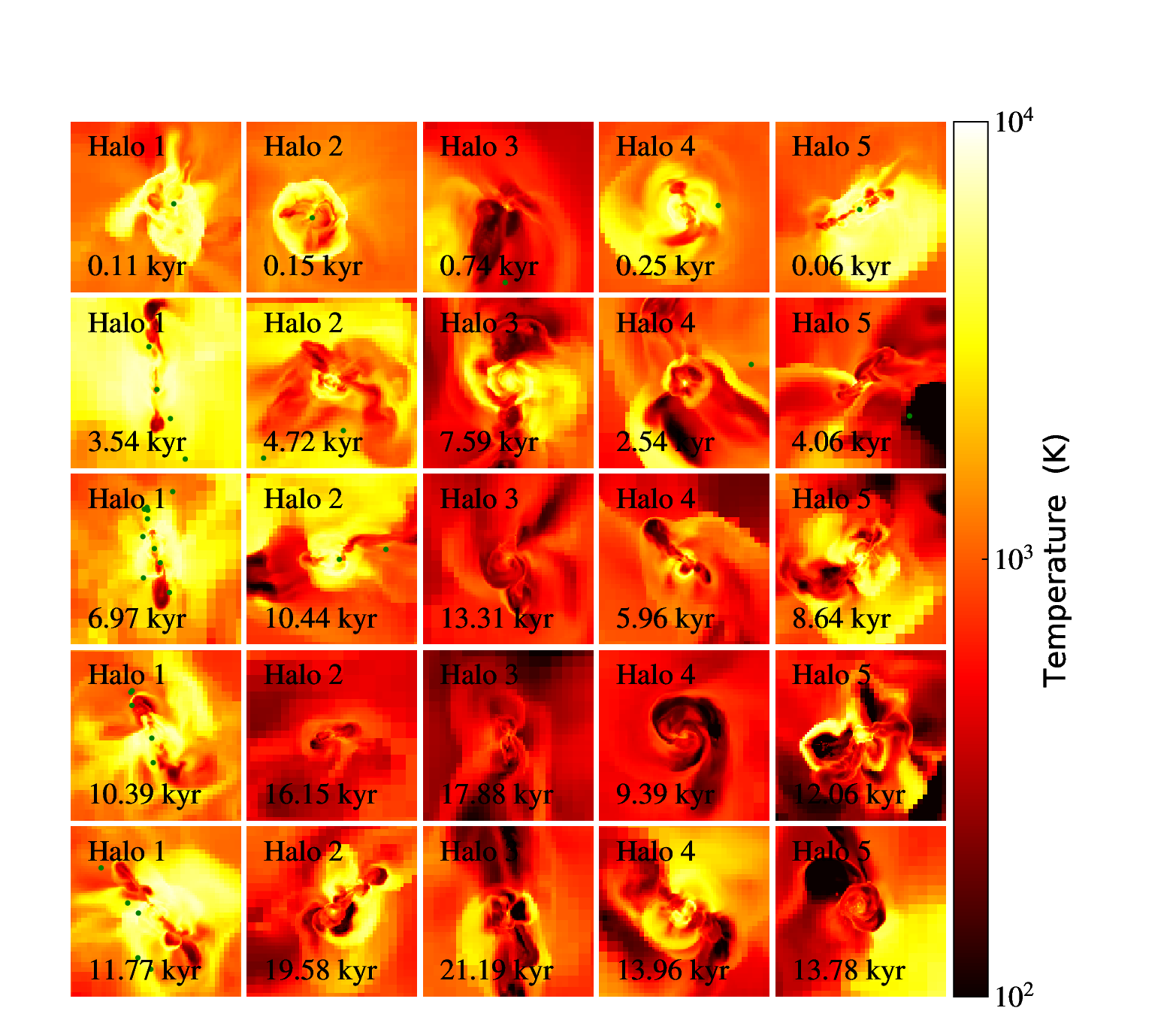} 
\end{center}
\vspace{-0.1cm}
\caption{Temperature slices along x-direction in the central 2000 AU of each halo. Green dots mark the positions of star particles.}
\label{fig:f4}
\end{figure*}

Density and temperature images of the halos are shown in Figures \ref{fig:f3} and \ref{fig:f4}.  The mass of the irst star to form in halo 1 is 20 \Ms\ and its radiation dissociates H$_2$ and ionizes and heats the gas to $\rm \sim10^4$ K, which generates a strong shock that drives the gas outwards (see Figure~\ref{fig:f3}).  The outflows have densities of $10^{-15} - 10^{-13}$ g cm$^{-3}$. About 920 yr later a second, 11 \Ms\ star forms in the disk. Radiation pressure from stellar flux compresses the disk and leads to the formation of four more stars over the next few hundred years.  Radiation from these stars breaks up the disk and creates dense clumps in which new stars form, and the number of stars rises to 10 over the next 4 kyr.  Subsequently, two short starbursts at 6 kyr and 8 kyr increase the number of stars to 16 and 21, respectively.  Stars continue to form in dense clumps over time and generate strong outflows, but they are unable to halt star formation because of infall from larger scales at rates of a few 10$^{-3}$ \Ms yr$^{-1}$. Mass inflow rates onto the disk are quite intermittent and mainly regulated by radiation from stars.

The first star in halo 2 is 25 \Ms\ and its ionizing UV flux launches a strong outflow that suppresses star formation in the disk for about 3.6 kyr.  The second star is 16 \Ms, and its radiation distorts the disk and forms an annular structure.  UV from both stars continues to quench star formation for the next 6 kyr and then six more stars form over the next 4.6 kyr with masses of 25 \Ms, 7 \Ms, 7 \Ms, 1.0 \Ms, 0.5 \Ms\ and 0.8 \Ms.  Radiation from these stars deforms the disk, suppresses star formation, and creates an annular shell of gas. In the final 8 kyr, only one subsolar-mass  object (0.001 \Ms) forms which may evolve into Pop III brown dwarf. Star formation drives a series of outflows that collide with infalling gas and are eventually absorbed by the surrounding dense medium at radii less than 3000 AU.

\begin{figure*}
\begin{center}
\begin{tabular}{cc}
\includegraphics[width=0.5\textwidth]{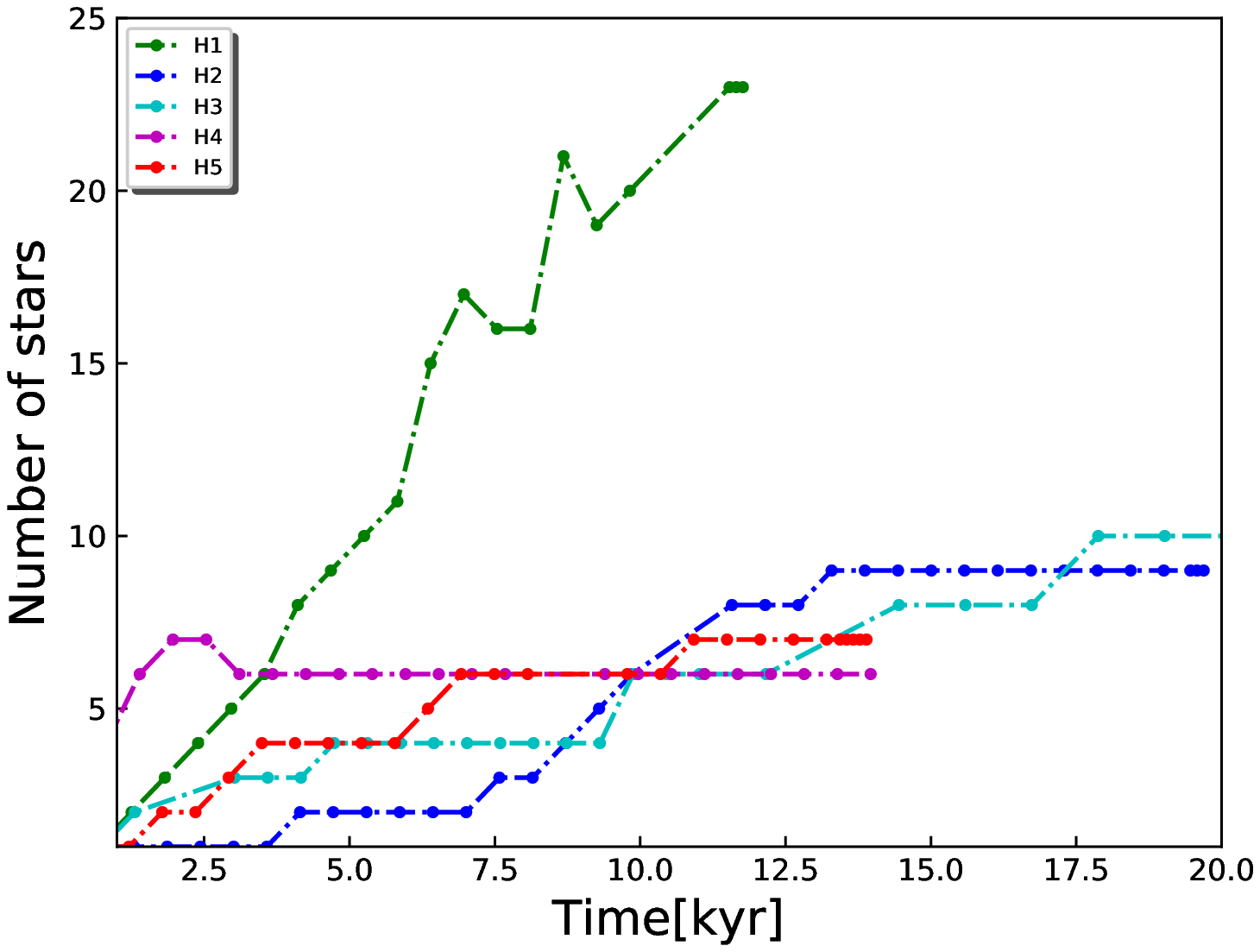} &
\includegraphics[width=0.5\textwidth]{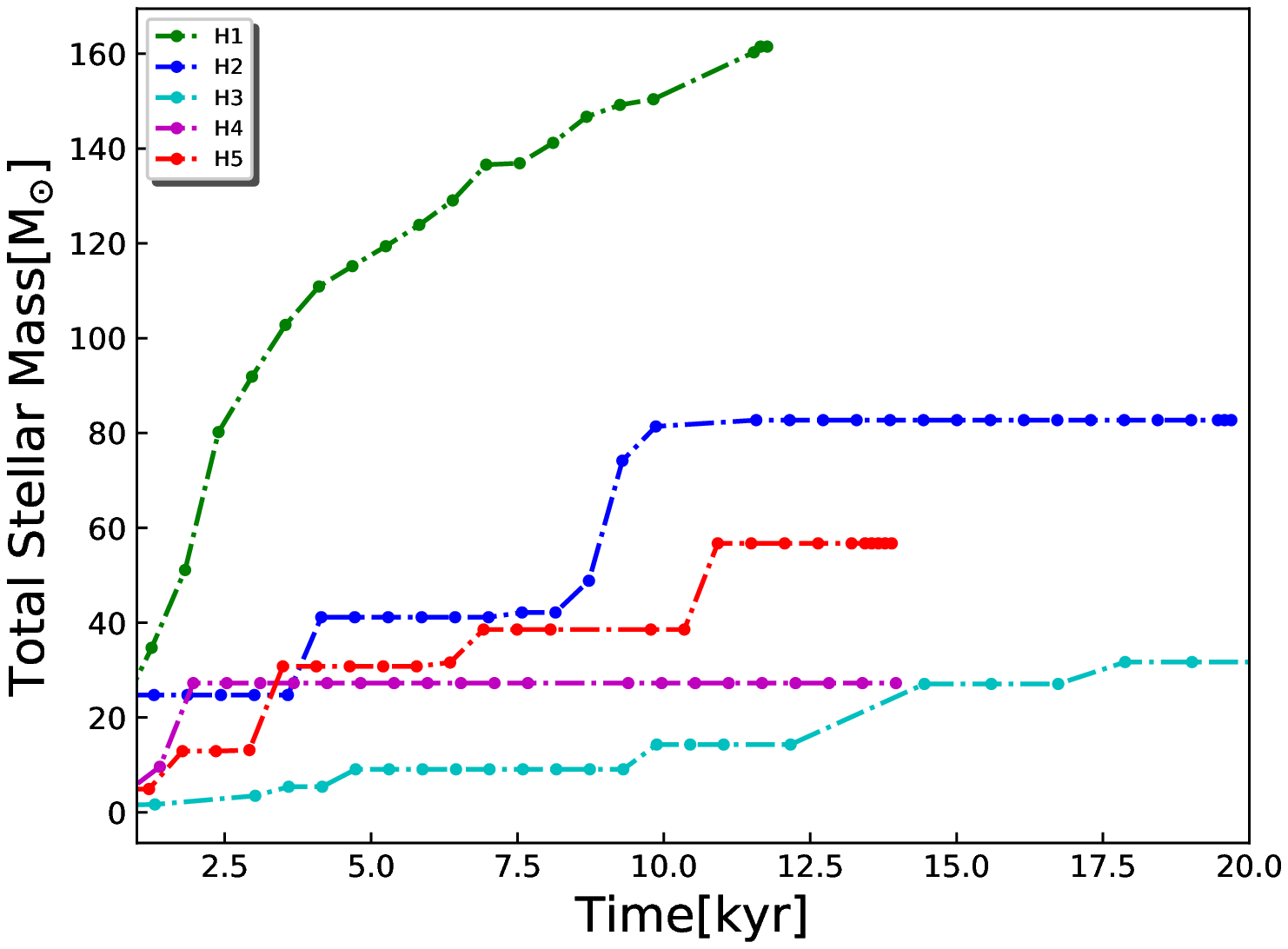} \\
\end{tabular}
\end{center}
\caption{Time evolution of number of stars (left panel) and stellar mass (right panel)  is shown here for all halos. Time is calculated after the formation of the first primary star in each halo.}
\label{fig:f5}
\end{figure*}

Most sink masses are subsolar or a few \Ms, but a few that are born in dense clumps are massive at birth. The mass of the first star in halo 3 is 2 \Ms\ and a second, 2 \Ms\ star forms after 790 yr.  The disk in this halo is quite compact and rotates at higher rates than the disks in halos 1 and 2, with initial mass accretion rates that are a factor of a few lower than in the other halos. Four stars form in halo 3 in the first 5 kyr compared to the two stars that form in this time in halo 2, but they have low masses ($< 5$ \Ms).  In total, 10 stars form over 12 kyr.  One is 13 \Ms\ and the others are 0.1 - 10 \Ms.  Star formation rates in halo 3 are similar to those in halo 2 but the stars are less massive.

A 3 $\rm \Ms$ star forms first in halo 4 and then six more stars appear over the next 2 kyr.  All are low-mass stars ($\rm < 5~ \Ms$) except one that is $\rm \sim18~ \Ms$, and two of the stars later merge.  Ionizing UV from the most massive star drives flows that sweep gas up into an expanding, ring-like structure. The ring fragments into multiple clumps, some of which later merge. Radiation from massive stars and dissipation heat from merging clumps suppress star formation for the last 10 kyr.  A total of 6 stars form in halo 4 over 13 kyr.

The first star that forms in halo 5 is less than a solar mass but it rapidly grows to 4.9 $\rm \Ms$  through accretion and a merger with another star that forms within its accretion radius. A second 8 $\rm \Ms$ star forms 1.3 kyr later.  Two more 0.9 \Ms\ and 17 \Ms\ stars appear over the next kyr, and ionizing UV from the more massive star plows up a ring-like structure that later fragments into multiple clumps.  After about 2.2 kyr, three more stars with masses of a few \Ms\ form over 4 kyr.  A massive 18 \Ms\ star then forms whose radiation blows away the surrounding gas and halts star formation for 34 kyr.  A total of 7 stars form in halo 5 over 14 kyr.  Outflows driven by the stars expand to a few thousand AU and collide with the dense infalling gas, which dissipates their energy.

We show the numbers of stars forming in the halos in the left panel of Figure~\ref{fig:f5}. In general, they increase time except in halo 4.  The total number of stars in halo 1 is 23:  three are $>$ 20 \Ms, five are $>$ 10 \Ms, eleven are 1 - 10 \Ms, and the rest are less than 1 \Ms. The dips in number of stars are due to the mergers with other stars. Nine stars form in halo 2 over 20 kyr:  two have masses above 20 \Ms, one is 16 \Ms, three are 1 - 10 \Ms, and three are less than 1 \Ms.  Ten stars form in halo 3 in 20 kyr:  nine are 0.1 - 7 \Ms\ and one is 13  \Ms.  Only six stars form in halo 4 over 14 kyr:  three are 1 - 5 \Ms, one is 18 \Ms\ and two are $<$ 1 \Ms.  In halo 5, 7 stars form over 14 kyr:  two are above 10 \Ms\ (18 and 17  \Ms), three are 4 - 8 \Ms, and two are $<$ 1 \Ms.

\begin{figure} 
\begin{center}
\includegraphics[scale=0.5]{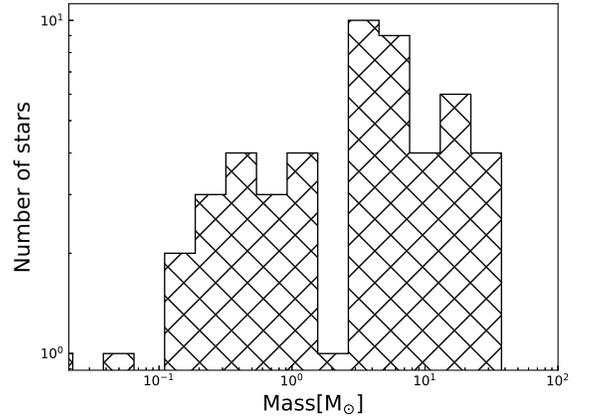} 
\end{center}
\vspace{-0.1cm}
\caption{Distribution of Pop III star masses in the five halos.}
\label{fig:f6}
\end{figure}

As shown in the right panel of Figure~\ref{fig:f5}, total stellar masses in the halos rise over time.  The bumps in the plots correspond to the formation of massive stars at later times and the plateaus are due to quiescent phases with no star formation. The final total stellar masses in halos 1 - 5 are 166 \Ms, 83 \Ms, 32 \Ms, 27 \Ms, and 57 \Ms, respectively.  Halo 1 has twice the average spin parameter of minihalos, and the higher angular momentum of the disk results in more fragmentation, as found in previous studies \citep{Latif20,pat21a}. Halo 2 has half the average minihalo spin parameter and average accretion rates that are a factor of two smaller than in halo 1. 

Higher disk rotation rates and lower accretion rates produce less massive stars in halo 3, below 10 $\rm \Ms$.  Only six stars form in halo 4 and only one is above 15 \Ms. Lower accretion rates, rapid disk rotation, collisional dissociation of H$_2$ from mergers between clumps, and radiative feedback by stars produce the small number of stars in halo 4.  Halos 4 and 5 have similar numbers of stars but the total stellar mass in halo 5 is higher because two of its stars are greater than15 \Ms.  The two halos were only evolved to 14 kyr so more stars may form at later times and the total stellar mass may increase. All in all, the differences in the total stellar masses are due to the complex interplay between halo spin, inflow rates onto the disks, and radiative feedback by massive stars.



We show the mass distribution of the stars in our ensemble of halos in Figure~\ref{fig:f6}.  Stellar masses range from  0.1 - 40 \Ms. Of the 55 stars, five are $>$ 20 \Ms, seven are 10 - 20 \Ms, 25 have masses of 1 - 10 \Ms, and 18 are less than 1 \Ms.  About 69\% of the total stellar mass is in the most massive stars ($\geq$ 10 \Ms),  suggesting a top-heavy initial mass function (IMF) with $dN/dM \propto M_*^{-1.2}$. This IMF is consistent with previous studies \citep{Susa14,Stacy16}.  Our results indicate that  about 70\% of the stars are ejected from their disks a few hundred years after birth. Initially, radiation from the primary stars dislocates the center of disk and three-body interactions with other stars later lead to their ejection.  The majority of the ejected stars have masses below 10 \Ms.  Ejection velocities of these stars are a few km s$^{-1}$, comparable to and in some cases even larger than the escape velocities of their host halos. This suggests that up to about $\sim$ 70\% of Pop III stars might be ejected from their host minihalos.

\begin{figure*}
\begin{center}
\begin{tabular}{cc}
\includegraphics[width=0.5\textwidth]{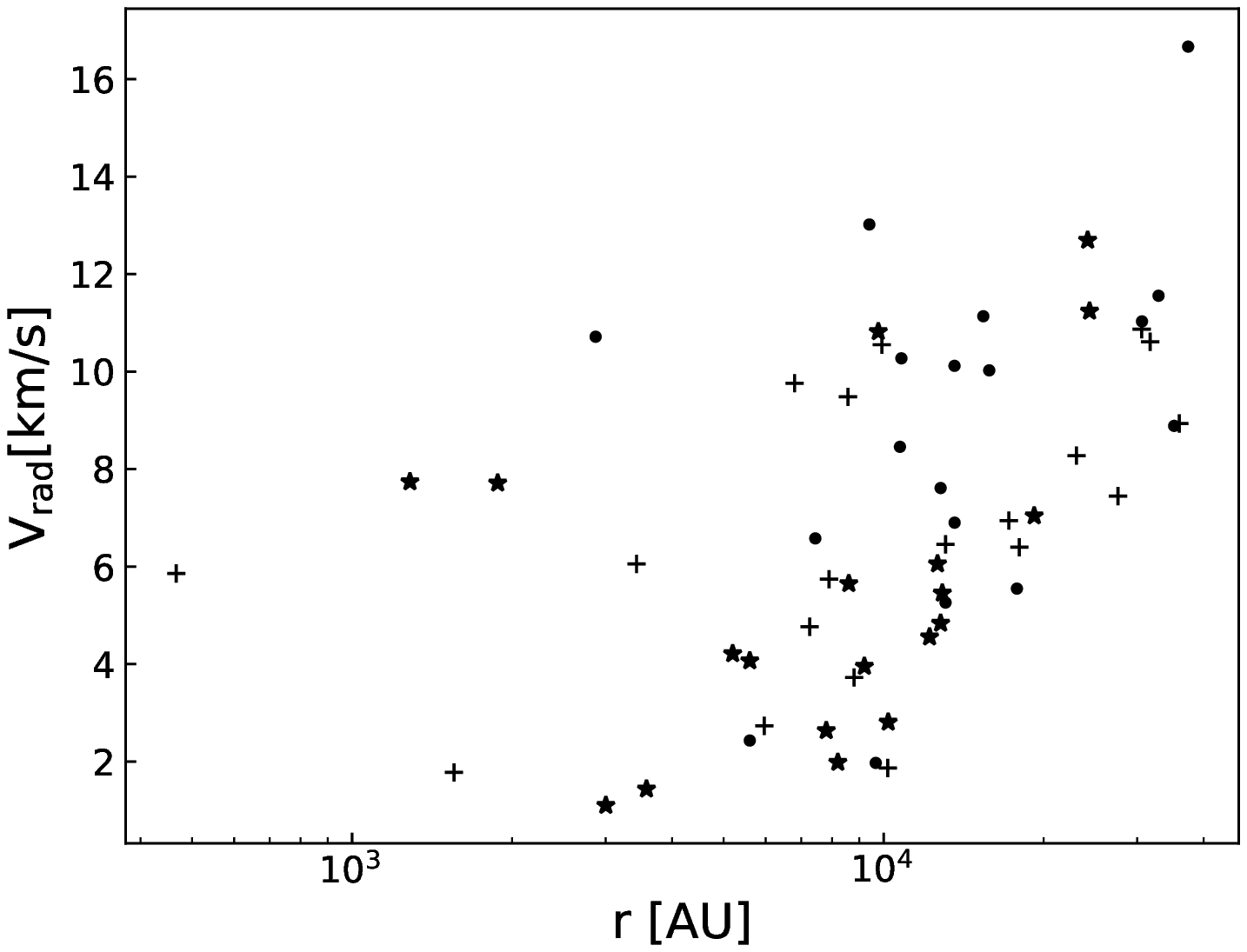} &
\includegraphics[width=0.5\textwidth]{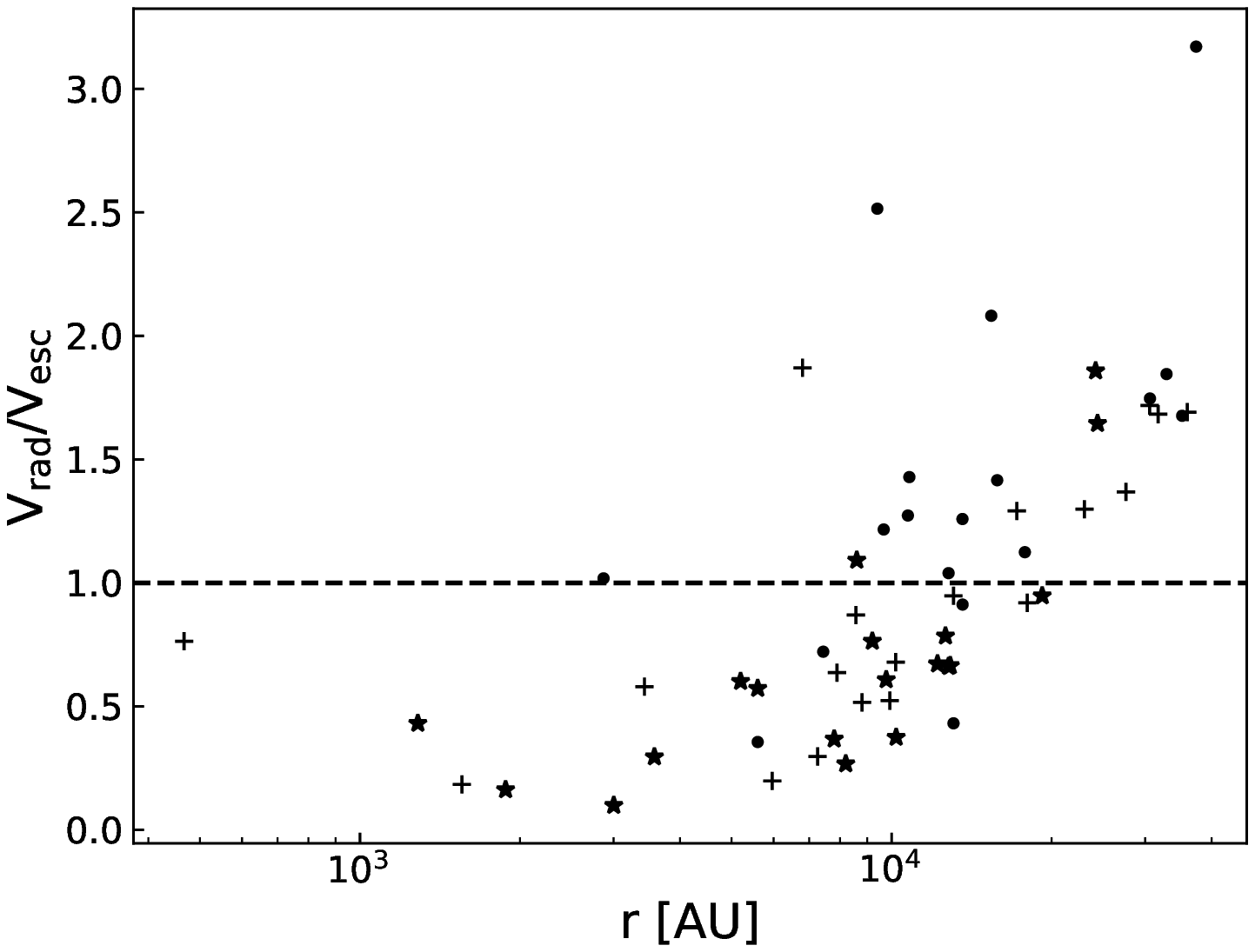} \\
\end{tabular}
\end{center}
\caption{Radial velocities (left) and the ratio of the radial velocity to the escape velocity of all stars (right panel) as a function of their distance from the center of mass. Black dots, crosses, and stars denote stars with masses below 1 \Ms, $ \rm 1 < M_{*} < 5$ \Ms, and above $ \rm 5 ~ \Ms$, respectively. The escape velocity is calculated from the total enclosed mass interior to the current radial distance of the star from the center of mass.}
\label{fig:f7}
\end{figure*}

Radial velocities and the ratio of the radial to escape velocities for all the stars is shown in Figure~\ref{fig:f7}. They have typical radial velocities of a few km s$^{-1}$, and most low-mass stars ($< 5$ \Ms) have $\rm v_{rad}/v_{esc}$ greater than 1. Low-mass stars are accelerated by dynamical interactions to velocities greater than $\rm v_{esc}$ and ejected. Our results are in agreement with \citep[][hereafter G11]{Greif11a}, who also find that low-mass protostars are ejected from the center of the cloud by dynamical interactions.  Most of the stars in our models are located a few thousand AU from the center of the center of the disk while in G11 they are a hundred AU from the center. This difference is due to the ionizing radiation in our simulations, which clears gas from the vicinity of the disk, and because our simulations are evolved for 20 times longer than in G11. Both factors contribute to their migration to larger radii.  Given these differences, our results are qualitatively similar to G11.

Current numerical schemes treat stars as point masses and cannot capture the collisions between physically extended protostars.  G11 approximated such collisions by relaxing the merging criteria of gravitational boundedness and allowing stellar mergers when they pass within 100 solar radii of each other, which led to more mergers in their simulations (so-called adhesive sink particles). Although we do not model the effect of gas-dynamical friction during stellar encounters, the accretion radius we use is larger than that in G11 and helps capture such mergers.  Nevertheless, the inclusion of such friction could result in more mergers and fewer ejections.

\section{Conclusion} \label{sec: discussion}



Our models suggest that ionizing UV from massive stars and ejections due to three-body interactions impose characteristic masses of 1-10 \Ms on Pop III stars.  Stars that are ejected from their protostellar disks are cut off from the accretion flows that form them and ionizing UV from the most massive stars in the disk limit accretion onto the others.  Up to 70\% of Pop III stars are ejected from their disks in our simulations, usually by a few hundred years after birth.  Most of the ejected stars have masses below 10 \Ms\ and velocities of a few km s$^{-1}$, comparable to and in a few cases greater than the escape velocities from their host halos.  The characteristic Pop III star masses in our simulation campaign are somewhat lower than those of previous studies because they did not include ionizing UV feedback from all the massive stars in the disk (and thus underestimated the effects radiative feedback) and did not evolve the disks for long enough times to tally ejections that terminated the growth of the stars.

We could not evolve the disks for longer than 20 kyr at resolutions of $\sim$ 10 AU because computational costs would have been prohibitive. Consequently, more stars may form at later times and some may even reach higher masses, but we expect the overall shape of the IMF to remain the same. We also could not fully resolve small-scale disks that may have formed around individual stars which could promote accretion and grow them to higher masses before the photoevaporation of the disk terminates their growth or they are ejected by three-body interactions. Fragmentation may also occur on smaller scales that are not resolved here.  However previous studies found that clumps forming on such scales in protostellar disks are expected to migrate inwards on short timescales and merge with the central star \citep{Latif15a,Hosokawa16}.

Our models do not capture all aspects of the pre-main sequence evolution of Pop III stars, but this does not strongly affect the dynamics of their outflows or the growth of the stars themselves. I-front radii mostly depend on ionizing photon emission rates, which we properly scale to stellar mass in our runs, and the expansion rate of the ionized flows depends on their temperatures, which are a weak function of the surface temperatures of the stars.  We therefore expect that our approximate treatment of the protostars will properly capture their early growth. Our simulations neglect magnetic fields, which are thought to form even in primordial protostellar disks at high redshifts via amplification by turbulent dynamos on small scales  \citep{Turk12, Schobera,ls15, Sharda20}.  Magnetic field lines in the disk may stabilize the disk against fragmentation and produce fewer, more massive stars.












\section{Acknowledgements}

MAL thanks the UAEU for funding via UPAR grant No. 31S390. %
\bibliography{smbhs}
\bibliographystyle{aasjournal}

\end{document}